\documentclass[%
 reprint,
 amsmath,amssymb,
 prl,
]{revtex4-2}
\usepackage{dcolumn}
\usepackage{bm}
\usepackage{hyperref}
\usepackage{comment}
\usepackage{mhchem}
\usepackage{xcolor}
\bibpunct[, ]{[}{]}{;}{n}{,}{,}
\usepackage{float}

\usepackage{color}
\usepackage{MnSymbol}
\definecolor{blue1}{rgb}{0.0,0.0,1.0}
\definecolor{blue2}{rgb}{0.0,0.1,1.0}
\definecolor{blue3}{rgb}{0.0,0.2,1.0}

\usepackage{lipsum,graphicx}

\renewcommand{\vec}[1]{\ensuremath{\mathbf{#1}}} 

\newcommand{\kBT}{k_\textsc{b} \,T}

\newcommand{\Nb}{N_\mathrm{b}}
\newcommand{\sigb}{\sigma_\mathrm{b}}
\newcommand{\taub}{\tau_\mathrm{b}}

\newcommand{\gb}{\gamma_\mathrm{b}}
\newcommand{\gr}{\gamma_\mathrm{r}}
\newcommand{\gs}{\gamma_\mathrm{s}}
\newcommand{\Pe}{\mathrm{Pe}}

\newcommand{\Fa}{F^\mathrm{a}}

\newcommand{\lmax}{l_\mathrm{max}}
\definecolor{barblue}{RGB}{153,204,254}
\definecolor{groupblue}{RGB}{51,102,254}
\definecolor{linkred}{RGB}{165,0,33}

\begin{document}
\preprint{APS/123-QED}
\title{Cluster and conquer: The morphodynamics of invasion of a compliant substrate by active rods}
\author{Mohammad Imaran$^{1,2,3}$, Mandar M Inamdar$^4$, Ranganathan Prabhakar$^2$, Raghunath Chelakkot$^3$}
\affiliation{$^1$IITB-Monash Research Academy, Mumbai, India\\
$^2$Department of Mechanical \& Aerospace Engineering, Monash University, Clayton, Australia\\
$^3$Department of Physics, Indian Institute of Technology Bombay, Mumbai, India\\
$^4$Department of Civil Engineering, Indian Institute of Technology Bombay, Mumbai, India\\
}

\date{\today}

\begin{abstract}
The colonisation of a soft passive material by motile cells such as bacteria is  common in biology. The resulting colonies of the invading cells are often observed to exhibit intricate patterns whose morphology and dynamics can depend on a number of factors, particularly the mechanical properties of the substrate and the motility of the individual cells.  We use  simulations of a minimal 2D model of self-propelled rods moving through with a passive compliant medium consisting of particles that offer elastic resistance before being plastically displaced from their equilibrium positions. It is observed that the motility-induced clustering of active (self-propelled) particles  is crucial for understanding the morphodynamics of colonisation. Clustering enables motile colonies to spread faster than they would have as isolated particles. The colonisation rate depends non-monotonically on substrate stiffness with a distinct maximum at a non-zero value of substrate stiffness. This is observed to be due to a change in the morphology of clusters.  Furrow networks created by the active particles have a fractal-like structure whose dimension varies systematically with substrate stiffness but is less sensitive to particle activity. The power-law growth exponent of the furrowed area is smaller than unity, suggesting that, to sustain such extensive furrow networks,  colonies must regulate their overall growth rate.
\end{abstract}

\maketitle



Large-scale collective pattern formation by self-motile elements is a widely studied phenomenon in physics and biology ~\cite{dombrowski2004self, ballerini2008interaction, katz2011inferring, marchetti2013hydrodynamics, ramaswamy2010mechanics}  . A prominent class of such collective behaviour such as flocking and swarming are caused by direct local interaction among individual entities within a group, influencing their relative motion \cite{vicsek1995novel, gregoire2004onset, chate2008modeling}. The surrounding medium can also play a significant part in mediating interactions between active particles. In wet systems, particularly at low Reynolds numbers, hydrodynamic interactions are nearly instantaneous.   Inhomogeneous and complex environments qualitatively change the individual, as well as collective dynamics of such systems~\cite{bechinger2016active,elgeti2015physics,qi2020enhanced,mousavi2020wall,D0SM00797H,patteson2018propagation,theeyancheri2020translational,das2020aggregate}. In dry systems, particles may leave behind chemical or mechanical cues for other particles to follow,  thereby increasing the level of medium-induced complexity. This phenomenon of stigmergy has been studied widely in the context of ants and termites leaving pheromone trails \cite{attygalle1985ant,jackson2004trail}.  

Recently, a mechanical form of stigmergy has been observed in motile bacteria \textit{Pseudomonas aeruginosa} and \textit{Myxococcus xanthus}  when cultured on soft hydrogel substrates ~\cite{gloag2013self,gloag2015bacterial,gloag2016stigmergy}. Under favourable conditions, these bacteria form extensive networks of permanent furrows as they move collectively as a monolayer across the soft surface in the initial stages of biofilm formation. The rate of expansion of the colony is further observed to be intimately related to the morphology of the network and the cellular traffic within the furrows~\cite{gloag2013self,gloag2015bacterial,gloag2016stigmergy}. 
 
Several recent studies have explored collective phenomena in populations of rod-shaped particles using computer simulations. The expansion rate and morphology of colonies of growing and sub-dividing non-motile have been shown to strongly depend on the mechanical properties of the passive medium or the extracellular material secreted by the rods ~\cite{farrell2013mechanically, ghosh2015mechanically,tchoufag2019mechanisms, you2018geometry, acemel2018computer,farrell2017mechanical, you2019mono}. Studies of self-propelled rods in the absence of coupled interactions with a substrate  \cite{peruani2006nonequilibrium,mccandlish2012spontaneous,abkenar2013collective,baskaran2008hydrodynamics,weitz2015self,prathyusha2018dynamically, velasco2018collective,kuan2015hysteresis,bar2020self,be2020phase,velasco2018collective,vliegenthart2020filamentous,bott2018isotropic} reveal self-organized formation of dynamic small or large clusters, polar lanes, and nematic defects.

\begin{figure*}[t]
\centerline{\includegraphics[width=0.8\linewidth]{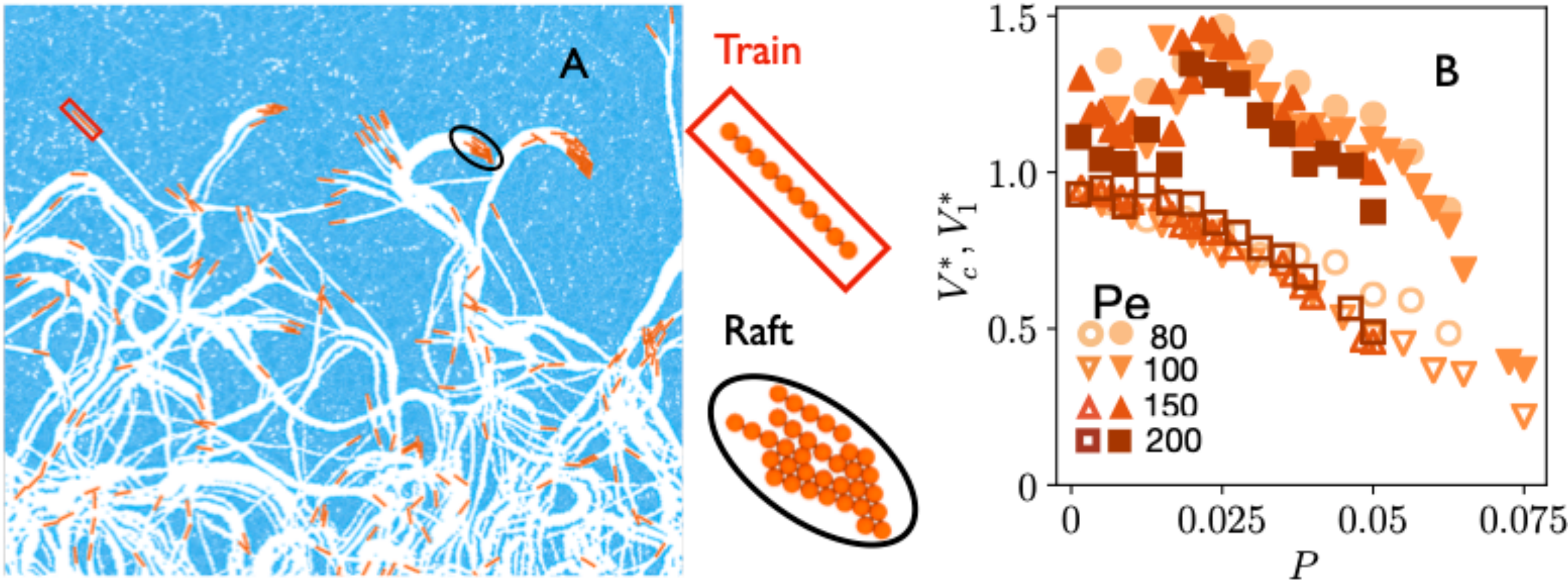}}
\vspace*{-1.3em}
\caption{ \label{fig:colonizn} (A) Furrow network (in white) formation in a substrate of plasticity, $P = 0.0025$,  at $\Pe = 100$. Active rods are red and substrate particles are blue. Adjacent figures show close-up images of two  distinct types of motility-induced clusters. 
Rafts are arrow-head shaped clusters, while trains consist of rods arranged end-to-end. (B) Comparison of normalised colonisation speed, $V_c^*$ (filled symbols), and the normalised speed of isolated rods, $V_1^*$ (open symbols). The speeds are normalised by the observed speed of a single rod in a substrate of zero stiffness.  When thermal fluctuations are weak, the average speed of an isolated rod in a fluid-like medium consisting of fully plasticised substrate is $V_1^0 \approx \Fa/ (\gs + \gr) = \Pe/ [\Nb^2\, (\gs/\gr + 1)]$.}
\label{fig1}
\vspace{-1.9em}
\end{figure*}

In an attempt to understand the behaviour observed in experiments in colonies of motile rod-shaped cells of \textit{P. aeruginosa} growing on agar \citep{gloag2013self}, Zachreson \textit{et al.} \citep{zachreson2017emergent,zachreson2017network} used simulations of self-propelled spherocylinders interacting with a continuum model of a deformable substrate. While their results  demonstrate that substrate stiffness and its viscoelastic relaxation time can strongly influence the morphodynamics of active furrowing, their parameters were chosen specifically for the experimental system at hand. Moreover, the simulations also included cell division and population growth. 

Here, we take a closer look by using 2D simulations to study the furrow structure that emerges as a single row of active self-propelled rods advances through a plastic substrate. We show that it is \textit{motility-induced clustering} that generically causes  furrow networks to emerge whose fractal dimension depend on substrate plasticity.  Clustering further enhances the rate at which the colony edge advances, and this speed gain depends on cluster morphology. Our results also suggest that colonies must regulate their overall cell growth rate in order to sustain extended furrow networks.

In our simulations, each rod of length $L \,= \, \Nb \, \sigb$ is  a rigid linear array of $\Nb$ beads of nominal diameter, $\sigma_b$. The substrate is discretised as randomly-packed isotropic particles (see Supplemental Information for detailed description \citep{SI}). The rods interact with other rods and substrate particles through repulsive excluded-volume forces modeled with the Separation-Shifted Lennards-Jones potential (SSLJ) \cite{abkenar2013collective}. The instantaneous configuration of rod $i$ are characterised by its position $\vec{r}_i(t)$ and the unit polarity vector $\hat{\vec{p}}_i(t)$. The time evolution of these variables are governed by Langevin equations that include random Brownian forces with an energy scale of $\kBT$ and a self-propulsion force of magnitude $\Fa$ that acts along $\hat{\vec{p}}_i$. Free rods experience frictional resistance to their motion characterised by the bare rod friction coefficient, $\gr \,=\, \gb \Nb$, where $\gb$ is the bare friction coefficient of a bead on a rod.  A minimal model is used to represent the essential mechanical features of a semi-solid, plastic substrate. The $p$-th substrate particle is bound to its equilibrium coordinate $\vec{r}_{p,\,0}$ by a harmonic potential, $U_p \,  =\,  k \,\left(\vec{r}_{p} -\vec{r}_{p,\,0}\right)^2/ 2$, if $ \mid\vec{r}_p -\vec{r}_{p,\,0}\mid \,< \lmax$, where $k$ is the elastic stiffness of the substrate, and $\lmax$ is the maximum displacement of substrate particles beyond which they are plasticised (\textit{i.e.} $U_p \,=\,k\,\lmax^2/2$, a constant) and a permanent furrow is formed (see Fig.~S1 \citep{SI}).
 Once broken off, the substrate particles remain as free particles in the system and continue to offer a frictional resistance to the rods. To model the high viscous resistance of semi-solid substrate to displacements,  the friction coefficient, of a substrate particle $\gs \,= \,10 \,\gb$ in our simulations. For the qualitative exploration here, we have neglected direct substrate-substrate interactions. The rod-substrate interaction forces and the binding force derived from the potential $U_p$ determine the time-evolution of the substrate. The simulations are performed using LAMMPS. The masses of the particles are chosen to be small enough for the simulations to be in the overdamped regime (see Supplemental Material \citep{SI}). 

Each simulation starts with two rows of vertically aligned rods, introduced at the top and bottom boundaries with propulsion forces directed towards the undisturbed substrate that fills rest of the box. The height of the simulation box , $H$, is twice its width, $W$. Periodic boundary conditions are imposed on all the sides. The simulation is stopped when any rod from either side crosses the middle of the box to the other side. Each side of the box is then treated as a separate simulation instance. We set  $\sigma_b$ as the length scale, and $\taub \,=\, \sigb^2 \gb / \kBT$ and $\kBT$ as the time and energy scales, respectively.  We use rods with $\Nb \, =\, 5$ in simulation boxes of dimensions $H \,=\, 400 \,\sigb$ and $W \,=\, 200 \,\sigb$.  The key dimensionless parameters are the P\'{e}clet number, $\Pe \, = \, \Fa\, L/ \kBT$ and the plasticity ratio, $ P \,=\, k \, \lmax/ \Fa$, with $\lmax \,=\, 0.6 \,\sigb$. We focus here on collective behaviour of the rods and its effect on furrow formation at high P\'{e}clet  numbers at which noise does not destroy structure formation. Our choice of parameters correspond to values of $P \ll 1$ \textit{i.e.} we consider plastic substrates for which the stiffness is chosen such that even single rods can displace the substrate to create permanent furrows.

\begin{figure*}[!t]
\centerline{ \includegraphics[width=0.95\textwidth]{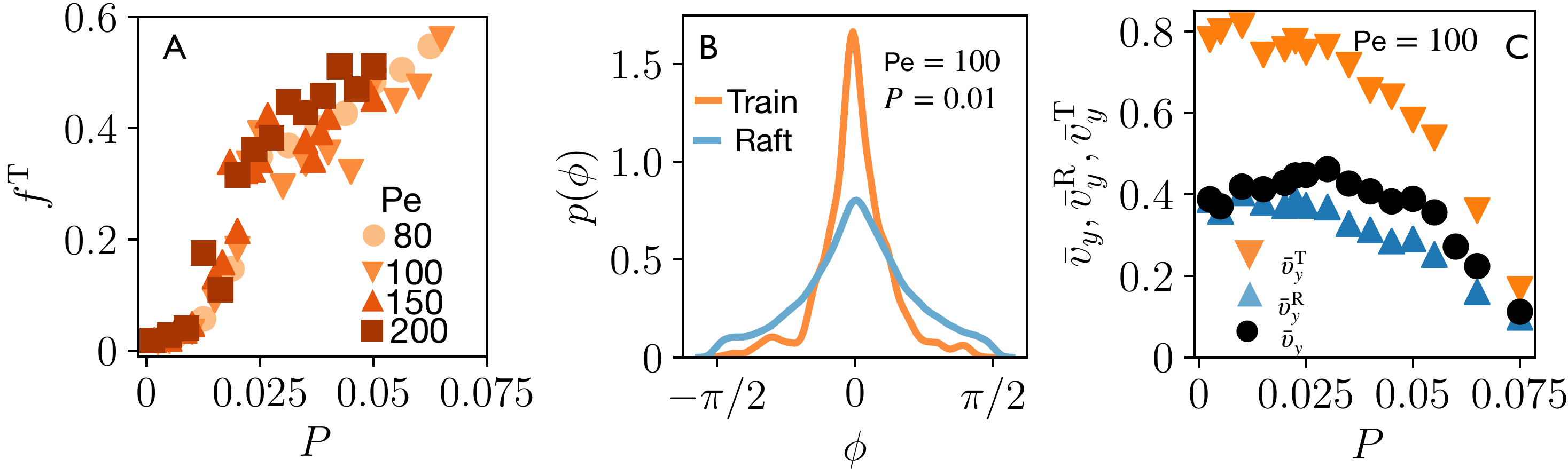}}
\vspace*{-1.4em}
\caption{\label{fig:stats}  (A) Fraction of train clusters, $f^\mathrm{T}$, out of all the  clusters ( the fraction of rafts, $f^\mathrm{R} \,= \,1  -  f^\mathrm{T}$). (B) Distribution of rod orientation angle $p(\phi)$ with respect to the vertical axis ($y$ axis) for rafts (blue curve) and trains (orange curve) for $\Pe \,=\,  100$ and $P \, = \, 0.01$. (C) Individual averages of the $y$-component of cluster velocity of rafts and trains, and the overall population average, for $\Pe\,=\,100$. }
\vspace{-2.0em}
\end{figure*}

Initially, all the rods begin individually creating relatively straight and narrow furrows (see Fig.~S3; Movie $\S$1, \citep{SI}). 
Small orientational fluctuations cause the rods to begin colliding with neighbouring rods to begin forming dynamic clusters. As the clusters propel through the substrate, they permanently displace substrate particles, forming a complex network of permanent furrows (Fig~\ref{fig1}-A). To analyse the furrow formation more quantitatively, we define the colonised area $C$, which is the $y$-coordinate of the outermost leader rod at any instant of time (see Fig.~S2 \citep{SI}). This area is observed to grow nearly linearly in time in all our simulations until they are terminated (see Fig.~S4 \citep{SI}).  The colonisation speed $V_c$ is estimated by a linear least-squares fit through the $C$-vs-$t$ data, and is compared in Fig.~\ref{fig:colonizn}-B with $V_1$, the speed of an isolated rod through the same substrate. The speeds are normalised by observed speed of a single rod in a substrate of zero stiffness.  We observe that, at  any given $\Pe$ and $P$, $V_c$ is systematically \textit{larger} than  $V_1$, the speed at which an isolated rod moves through the substrate. Further, while $V_1$ decreases nearly linearly with $P$ at a fixed $\Pe$, $V_c$ varies non-monotonically with $P$, displaying a peak colonisation rate at a non-zero value of $P$. 

These interesting differences in colonisation speed from isolated rod speed clearly arise from collective effects and appear to be related to  the behaviour of  clusters of active rods that drive the formation of furrows.  We used a clustering algorithm to identify separate clusters of beads. Individual clusters were then tracked in a frontier region of depth $8 \,L$ from the leading edge of the colony until they broke up or were joined by new members. Frequency distributions of quantities such as the number of rods in a cluster (\textit{i.e.} the cluster size), average cluster speed and vertical velocity component, and orientation angle relative to the vertical axis were  determined. 

Based on the relative configurations of the rods in a cluster, each cluster is categorised as one of two mutually exclusive types: end-to-end \textit{trains} and non-train \textit{rafts} that are usually arrow-head shaped (Fig.~\ref{fig1}-A; see Supplementary Material for mathematical definitions \citep{SI}). 
We find that trains become  more prevalent as the substrate becomes stiffer (Fig.~\ref{fig:stats}-A;  Movie $\S$2, \citep{SI}). 

Both rafts and trains move more quickly through the substrate than isolated rods. Since the propulsive force on each rod is distributed uniformly over each bead, the total propulsive force on a tightly-packed  cluster is proportional to its area. The resistance it experiences is however proportional  to its frontal perimeter exposed to the substrate. Consequently, the resistance experienced per rod  is smaller in a cluster. We therefore find that the average speed of clusters of either kind grows linearly with cluster size (Fig.~S5 in SI \citep{SI}). This effect is greater in trains, where the whole propulsive force of the train is only resisted by about one or two substrate particles.  

In addition, the orientational distribution of clusters shows that trains in the frontier region close to leading edge of the colony tend to be strongly aligned in the outward direction (Figs.~\ref{fig:stats}-B and S7 in SI \citep{SI}).  Rods in rafts have more diverse orientations, which can cause rafts to have a broader range of orientations relative to the outward direction.  The propulsive force on trains, on the other hand, is almost entirely directed along the train axis, which keeps them on course.

The mean $y$-component of the velocity ($v_y$) of clusters  within the frontier region is computed as $\overline{v}_y \,= \,\overline{v}^\mathrm{R}_{y}\,( 1 - f^\mathrm{T}) \,+\,\overline{v}^\mathrm{T}_{y}\, f^\mathrm{T}$, where $f^\mathrm{T}$ is the fraction of trains in the  clusters.
In Fig.~\ref{fig:stats}-C, the individual means, $\overline{v}^\mathrm{R}_{y}$ and $\overline{v}^\mathrm{T}_{y}$, decrease non-monotonically with $P$ , the overall average $\overline{v}_y$ shows a maximum around the same $P$ at which the peak colonisation speed occurs in Fig.~\ref{fig:colonizn}-B. Therefore,  the increase of the colonisation speed with stiffness at small values of  $P$, appears to be because of the growth in the fraction of faster-moving trains that are also more aligned along the outward direction. When the fraction of trains begins to saturate at high stiffnesses, the decrease of their speeds with stiffness takes over, and the overall colonisation speed decreases with stiffness. 

The furrow network is created by clusters at the head of furrows repeatedly intersecting with  previously created, empty, furrows. If a previous furrow  a cluster encounters is narrow, the cluster can plough through and continue unhindered.  On the other hand, when the previous furrow has a width comparable to $L$ or greater, the cluster quickly breaks up, as the small orientational noise present causes rods at the edge of the cluster to escape the cluster easily before the cluster passes through the furrow (see Movie $\S$3, \citep{SI}). This is consistent with observations elsewhere that motility-induced clustering can lead to large clusters and lanes in systems of free active rods at sufficiently high densities  \citep{abkenar2013collective}, but in an empty furrow, a cluster can quickly ``evaporate''. 
We observe visually that arrow-head shaped rafts are stable for longer times. The rods at the head of such pointy rafts have convergent orientations. This leads to a rectification of their propulsion forces along the longitudinal axis of the cluster. Any remaining transverse component of the net propulsion force causes pointy rafts to swerve and take curved trajectories. Large  rafts can thus create wide and long furrows that serve as arterial conduits in the network  (Fig.~S8-D; Movie $\S7$ \citep{SI}). Broad-headed rafts with rods of divergent orientations quickly break up into multiple smaller rafts and corresponding furrows.

Collisions of free rods with existing clusters at the head of furrows play an important role in the process of network formation (Fig.~S8 A--C \citep{SI}). Free rods in empty furrows experience lower (bare) friction and move through the network very quickly.  These rods sometimes encounter a furrow wall head-on and can push through to create thin furrows on their own. However, more typically, on colliding with furrow walls, they reorient along the furrow axis and tend to catch up with one of the other rafts at the head of a newly forming furrow (or exit the periodic boundary at the back of their colony to enter the colony on the other side and catch up with a cluster at a furrow head on that side). Free rods approaching a raft or a train from behind cannot overtake the cluster. They either collide with the cluster from behind, or may, at best, squeeze through the side of the furrow to align with other rods at the head (Fig.~S8-A and B; Movies $\S4$ and $\S5$  \citep{SI}). These frequent collisions change the nematic alignment of the rods in the cluster.  Collisions thus lead to the break-up of clusters and the formation of new ones (see Fig.~S8-C; Movie $\S$6 \citep{SI}).

\begin{figure}
\centerline{\includegraphics[width=\columnwidth]{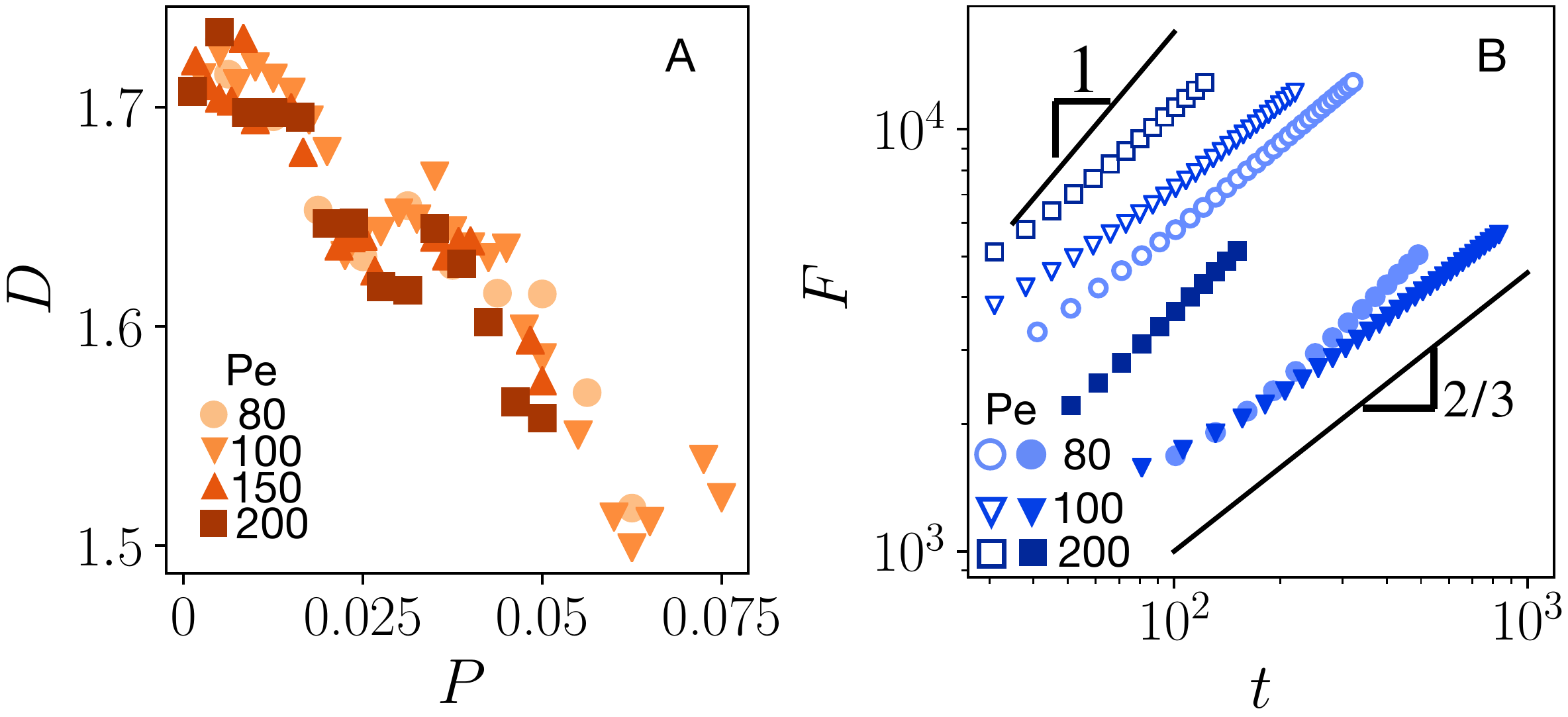}}
\vspace*{-1.4em}
\caption{\label{fig:clusters}  (A) Box-counting fractal dimension as a function of plasticity number $P$, measured at termination of simulation.  (B) The growth of total furrowed area with time, for different $Pe$ and $P$. furrowed area growth shows a non-trivial power law behaviour. The symbols are as follows: $P = 0.00125$( \textcolor{blue1}{$\medsquare$}) and $P = 0.05$(\textcolor{blue1}{$\filledmedsquare$}) at $\Pe =200$; $P = 0.0025$(\textcolor{blue2}{$\medtriangledown$}) and $P = 0.075$ (\textcolor{blue2}{$\filledmedtriangledown$}) at $\Pe=100$;  $P=0.00625$ (\textcolor{blue3}{$\medcircle$}) and $P=0.0625$(\textcolor{blue3}{ \Large $\bullet$}) at $\Pe=80$.}  
\vspace{-2.0em}
\end{figure}


The trajectories and collisions of rafts and trains leads to a distinctive network morphology. The highly-ramified structure of the networks are quantified here by determining the fractal dimension, $D$, of the furrow-substrate interface using a box-counting algorithm \cite{gagnepain1986fractal}. As expected, $D$ has a non-integer value between 1 and 2, and is observed to systematically decrease with  $P$ (Fig.~\ref{fig:clusters}-A) . This is associated with a change from networks created mostly by interactions of curved raft trajectories (\emph{e.g.} Fig.~\ref{fig:colonizn}-A) to  those dominated by the thin and straight furrows created by trains (\emph{e.g.} Fig.~\ref{fig:colonizn}-A)  

This observation suggests a simple model for the growth of the area of the furrow network, $F$, with time. Furrow heads can form randomly at any point on the furrow interface within the network due to the constant scattering of free rods through the network by cluster breakup. The furrowing rate is then expected to be proportional to the length, $Z$, of the furrow interface. The fractal nature of the interface implies that $Z \sim F^{D/2}$  \cite{florio2019use}.
The rate of furrow growth, $d F/ dt\, \sim\, Z\, R/\mathcal{F}$, where $R$ is the total area occupied by rods in the furrows (a constant in our simulations). 
Integrating, we obtain 
$ F \sim  t^{ 2/ (4-D)}$ at large times. Since the fractal dimension, $1 < D < 2$,  we observe in Fig.~ \ref{fig:clusters}-B that the power-law exponent for the growth of  $F$ with $t$  is in the range  $2/3 < 2/ (4-D) < 1$. If this can be sustained, while the colonized area $C$ grows linearly with time, the colony morphology will become increasingly sparse.

Based on the results above, one can expect similar furrow networks to generically form when colonies of motile cells  spread through soft, plastic environments and in three dimensions as well.  The observations here are in line with  many experiments with motile bacteria. Cell rafts similar to those discussed above have been observed to lead the formation of furrow networks in  \textit{P. aeruginosa} and \textit{M. xanthus} \citep{gloag2013self,  gloag2016stigmergy} cultured under confinement on semi-solid agar. These furrow networks extend across distances around two orders of magnitude greater than the length of a single cell. The networks are  further lined with extracellular DNA (eDNA) and extracellular polysaccharides (EPS). Other experiments have further demonstrated that the three-dimensional lattices of eDNA are necessary for the mechanical integrity of  biofilms of bacterial pathogens  in complex biofluids such as sputum and otorrhea (ear-discharge fluid)\citep{Devaraj2019}. The current study provides a natural mechanism for such network templates to spontaneously emerge in such systems.

Our observation of the power-law growth of $F$ further suggests that, in order to create extended furrow networks such as those observed in experiments,  there must be mechanisms that limit the growth rate of cells to not exceed the rate at which the furrows are created.  Unconstrained exponential growth or even linear growth of cells quickly obliterates the furrow network, leading to a dense colony (Fig.~S9; Movie $\S$8 \citep{SI}).
 Thereafter, the edge of the colony advances not by motile clusters, but by  bulk motion of a densely-packed front. While this can lead to fingering patterns at the front \citep{Nagilla2018,giverso2015branching}, a furrow network cannot be sustained. Explosive cell lysis events have, in fact, been observed in furrow networks of \textit{P. aeruginosa} along with associated release of eDNA and other ``public goods'' required for biofilm formation  \citep{Turnbull2016}. Beyond bacterial collectives, substrate interactions also occur during cancer metastasis in which groups of migrating cancer cells exploit the  plasticity of the extracellular matrix (ECM) and actively remodel the ECM fibers to enhance their migration speeds\cite{lee2017local,wisdom2018matrix, winkler2020concepts}.  

In conclusion, the cluster-and-conquer mechanism observed in our simulations may be crucial in biofilm formation by motile bacteria in plastic environments, and experimental evidence suggests that bacteria may have adaptive mechanisms to exploit the fractal-like spatial structure of the furrow network in building robust biofilms.  The results here suggest future theoretical and experimental avenues for exploring the role played by mechanical stigmergy in these biological phenomena.

\textit{Acknowledgments}: RC acknowledges the financial support from SERB, India via the grants SB/S2/RJN-051/2015 and ECR/2017/000744. We  acknowledge financial support from IITB-Monash Research Academy and  computational-time grants from the National Computational Infrastructure, Canberra, the MonArch facility at Monash University and the SpaceTime facility (IIT Bombay).



\begin{thebibliography}{52}%
\makeatletter
\providecommand \@ifxundefined [1]{%
 \@ifx{#1\undefined}
}%
\providecommand \@ifnum [1]{%
 \ifnum #1\expandafter \@firstoftwo
 \else \expandafter \@secondoftwo
 \fi
}%
\providecommand \@ifx [1]{%
 \ifx #1\expandafter \@firstoftwo
 \else \expandafter \@secondoftwo
 \fi
}%
\providecommand \natexlab [1]{#1}%
\providecommand \enquote  [1]{``#1''}%
\providecommand \bibnamefont  [1]{#1}%
\providecommand \bibfnamefont [1]{#1}%
\providecommand \citenamefont [1]{#1}%
\providecommand \href@noop [0]{\@secondoftwo}%
\providecommand \href [0]{\begingroup \@sanitize@url \@href}%
\providecommand \@href[1]{\@@startlink{#1}\@@href}%
\providecommand \@@href[1]{\endgroup#1\@@endlink}%
\providecommand \@sanitize@url [0]{\catcode `\\12\catcode `\$12\catcode
  `\&12\catcode `\#12\catcode `\^12\catcode `\_12\catcode `\%12\relax}%
\providecommand \@@startlink[1]{}%
\providecommand \@@endlink[0]{}%
\providecommand \url  [0]{\begingroup\@sanitize@url \@url }%
\providecommand \@url [1]{\endgroup\@href {#1}{\urlprefix }}%
\providecommand \urlprefix  [0]{URL }%
\providecommand \Eprint [0]{\href }%
\providecommand \doibase [0]{https://doi.org/}%
\providecommand \selectlanguage [0]{\@gobble}%
\providecommand \bibinfo  [0]{\@secondoftwo}%
\providecommand \bibfield  [0]{\@secondoftwo}%
\providecommand \translation [1]{[#1]}%
\providecommand \BibitemOpen [0]{}%
\providecommand \bibitemStop [0]{}%
\providecommand \bibitemNoStop [0]{.\EOS\space}%
\providecommand \EOS [0]{\spacefactor3000\relax}%
\providecommand \BibitemShut  [1]{\csname bibitem#1\endcsname}%
\let\auto@bib@innerbib\@empty
\bibitem [{\citenamefont {Dombrowski}\ \emph {et~al.}(2004)\citenamefont
  {Dombrowski}, \citenamefont {Cisneros}, \citenamefont {Chatkaew},
  \citenamefont {Goldstein},\ and\ \citenamefont
  {Kessler}}]{dombrowski2004self}%
  \BibitemOpen
  \bibfield  {author} {\bibinfo {author} {\bibfnamefont {C.}~\bibnamefont
  {Dombrowski}}, \bibinfo {author} {\bibfnamefont {L.}~\bibnamefont
  {Cisneros}}, \bibinfo {author} {\bibfnamefont {S.}~\bibnamefont {Chatkaew}},
  \bibinfo {author} {\bibfnamefont {R.~E.}\ \bibnamefont {Goldstein}},\ and\
  \bibinfo {author} {\bibfnamefont {J.~O.}\ \bibnamefont {Kessler}},\
  }\href@noop {} {\bibfield  {journal} {\bibinfo  {journal} {Physical review
  letters}\ }\textbf {\bibinfo {volume} {93}},\ \bibinfo {pages} {098103}
  (\bibinfo {year} {2004})}\BibitemShut {NoStop}%
\bibitem [{\citenamefont {Ballerini}\ \emph {et~al.}(2008)\citenamefont
  {Ballerini}, \citenamefont {Cabibbo}, \citenamefont {Candelier},
  \citenamefont {Cavagna}, \citenamefont {Cisbani}, \citenamefont {Giardina},
  \citenamefont {Lecomte}, \citenamefont {Orlandi}, \citenamefont {Parisi},
  \citenamefont {Procaccini} \emph {et~al.}}]{ballerini2008interaction}%
  \BibitemOpen
  \bibfield  {author} {\bibinfo {author} {\bibfnamefont {M.}~\bibnamefont
  {Ballerini}}, \bibinfo {author} {\bibfnamefont {N.}~\bibnamefont {Cabibbo}},
  \bibinfo {author} {\bibfnamefont {R.}~\bibnamefont {Candelier}}, \bibinfo
  {author} {\bibfnamefont {A.}~\bibnamefont {Cavagna}}, \bibinfo {author}
  {\bibfnamefont {E.}~\bibnamefont {Cisbani}}, \bibinfo {author} {\bibfnamefont
  {I.}~\bibnamefont {Giardina}}, \bibinfo {author} {\bibfnamefont
  {V.}~\bibnamefont {Lecomte}}, \bibinfo {author} {\bibfnamefont
  {A.}~\bibnamefont {Orlandi}}, \bibinfo {author} {\bibfnamefont
  {G.}~\bibnamefont {Parisi}}, \bibinfo {author} {\bibfnamefont
  {A.}~\bibnamefont {Procaccini}}, \emph {et~al.},\ }\href@noop {} {\bibfield
  {journal} {\bibinfo  {journal} {Proceedings of the national academy of
  sciences}\ }\textbf {\bibinfo {volume} {105}},\ \bibinfo {pages} {1232}
  (\bibinfo {year} {2008})}\BibitemShut {NoStop}%
\bibitem [{\citenamefont {Katz}\ \emph {et~al.}(2011)\citenamefont {Katz},
  \citenamefont {Tunstr{\o}m}, \citenamefont {Ioannou}, \citenamefont {Huepe},\
  and\ \citenamefont {Couzin}}]{katz2011inferring}%
  \BibitemOpen
  \bibfield  {author} {\bibinfo {author} {\bibfnamefont {Y.}~\bibnamefont
  {Katz}}, \bibinfo {author} {\bibfnamefont {K.}~\bibnamefont {Tunstr{\o}m}},
  \bibinfo {author} {\bibfnamefont {C.~C.}\ \bibnamefont {Ioannou}}, \bibinfo
  {author} {\bibfnamefont {C.}~\bibnamefont {Huepe}},\ and\ \bibinfo {author}
  {\bibfnamefont {I.~D.}\ \bibnamefont {Couzin}},\ }\href@noop {} {\bibfield
  {journal} {\bibinfo  {journal} {Proceedings of the National Academy of
  Sciences}\ }\textbf {\bibinfo {volume} {108}},\ \bibinfo {pages} {18720}
  (\bibinfo {year} {2011})}\BibitemShut {NoStop}%
\bibitem [{\citenamefont {Marchetti}\ \emph {et~al.}(2013)\citenamefont
  {Marchetti}, \citenamefont {Joanny}, \citenamefont {Ramaswamy}, \citenamefont
  {Liverpool}, \citenamefont {Prost}, \citenamefont {Rao},\ and\ \citenamefont
  {Simha}}]{marchetti2013hydrodynamics}%
  \BibitemOpen
  \bibfield  {author} {\bibinfo {author} {\bibfnamefont {M.~C.}\ \bibnamefont
  {Marchetti}}, \bibinfo {author} {\bibfnamefont {J.-F.}\ \bibnamefont
  {Joanny}}, \bibinfo {author} {\bibfnamefont {S.}~\bibnamefont {Ramaswamy}},
  \bibinfo {author} {\bibfnamefont {T.~B.}\ \bibnamefont {Liverpool}}, \bibinfo
  {author} {\bibfnamefont {J.}~\bibnamefont {Prost}}, \bibinfo {author}
  {\bibfnamefont {M.}~\bibnamefont {Rao}},\ and\ \bibinfo {author}
  {\bibfnamefont {R.~A.}\ \bibnamefont {Simha}},\ }\href@noop {} {\bibfield
  {journal} {\bibinfo  {journal} {Reviews of Modern Physics}\ }\textbf
  {\bibinfo {volume} {85}},\ \bibinfo {pages} {1143} (\bibinfo {year}
  {2013})}\BibitemShut {NoStop}%
\bibitem [{\citenamefont {Ramaswamy}(2010)}]{ramaswamy2010mechanics}%
  \BibitemOpen
  \bibfield  {author} {\bibinfo {author} {\bibfnamefont {S.}~\bibnamefont
  {Ramaswamy}},\ }\href@noop {} {\bibfield  {journal} {\bibinfo  {journal}
  {Annu. Rev. Condens. Matter Phys.}\ }\textbf {\bibinfo {volume} {1}},\
  \bibinfo {pages} {323} (\bibinfo {year} {2010})}\BibitemShut {NoStop}%
\bibitem [{\citenamefont {Vicsek}\ \emph {et~al.}(1995)\citenamefont {Vicsek},
  \citenamefont {Czir{\'o}k}, \citenamefont {Ben-Jacob}, \citenamefont
  {Cohen},\ and\ \citenamefont {Shochet}}]{vicsek1995novel}%
  \BibitemOpen
  \bibfield  {author} {\bibinfo {author} {\bibfnamefont {T.}~\bibnamefont
  {Vicsek}}, \bibinfo {author} {\bibfnamefont {A.}~\bibnamefont {Czir{\'o}k}},
  \bibinfo {author} {\bibfnamefont {E.}~\bibnamefont {Ben-Jacob}}, \bibinfo
  {author} {\bibfnamefont {I.}~\bibnamefont {Cohen}},\ and\ \bibinfo {author}
  {\bibfnamefont {O.}~\bibnamefont {Shochet}},\ }\href@noop {} {\bibfield
  {journal} {\bibinfo  {journal} {Physical review letters}\ }\textbf {\bibinfo
  {volume} {75}},\ \bibinfo {pages} {1226} (\bibinfo {year}
  {1995})}\BibitemShut {NoStop}%
\bibitem [{\citenamefont {Gr{\'e}goire}\ and\ \citenamefont
  {Chat{\'e}}(2004)}]{gregoire2004onset}%
  \BibitemOpen
  \bibfield  {author} {\bibinfo {author} {\bibfnamefont {G.}~\bibnamefont
  {Gr{\'e}goire}}\ and\ \bibinfo {author} {\bibfnamefont {H.}~\bibnamefont
  {Chat{\'e}}},\ }\href@noop {} {\bibfield  {journal} {\bibinfo  {journal}
  {Physical review letters}\ }\textbf {\bibinfo {volume} {92}},\ \bibinfo
  {pages} {025702} (\bibinfo {year} {2004})}\BibitemShut {NoStop}%
\bibitem [{\citenamefont {Chat{\'e}}\ \emph {et~al.}(2008)\citenamefont
  {Chat{\'e}}, \citenamefont {Ginelli}, \citenamefont {Gr{\'e}goire},
  \citenamefont {Peruani},\ and\ \citenamefont {Raynaud}}]{chate2008modeling}%
  \BibitemOpen
  \bibfield  {author} {\bibinfo {author} {\bibfnamefont {H.}~\bibnamefont
  {Chat{\'e}}}, \bibinfo {author} {\bibfnamefont {F.}~\bibnamefont {Ginelli}},
  \bibinfo {author} {\bibfnamefont {G.}~\bibnamefont {Gr{\'e}goire}}, \bibinfo
  {author} {\bibfnamefont {F.}~\bibnamefont {Peruani}},\ and\ \bibinfo {author}
  {\bibfnamefont {F.}~\bibnamefont {Raynaud}},\ }\href@noop {} {\bibfield
  {journal} {\bibinfo  {journal} {The European Physical Journal B}\ }\textbf
  {\bibinfo {volume} {64}},\ \bibinfo {pages} {451} (\bibinfo {year}
  {2008})}\BibitemShut {NoStop}%
\bibitem [{\citenamefont {Bechinger}\ \emph {et~al.}(2016)\citenamefont
  {Bechinger}, \citenamefont {Di~Leonardo}, \citenamefont {L{\"o}wen},
  \citenamefont {Reichhardt}, \citenamefont {Volpe},\ and\ \citenamefont
  {Volpe}}]{bechinger2016active}%
  \BibitemOpen
  \bibfield  {author} {\bibinfo {author} {\bibfnamefont {C.}~\bibnamefont
  {Bechinger}}, \bibinfo {author} {\bibfnamefont {R.}~\bibnamefont
  {Di~Leonardo}}, \bibinfo {author} {\bibfnamefont {H.}~\bibnamefont
  {L{\"o}wen}}, \bibinfo {author} {\bibfnamefont {C.}~\bibnamefont
  {Reichhardt}}, \bibinfo {author} {\bibfnamefont {G.}~\bibnamefont {Volpe}},\
  and\ \bibinfo {author} {\bibfnamefont {G.}~\bibnamefont {Volpe}},\
  }\href@noop {} {\bibfield  {journal} {\bibinfo  {journal} {Reviews of Modern
  Physics}\ }\textbf {\bibinfo {volume} {88}},\ \bibinfo {pages} {045006}
  (\bibinfo {year} {2016})}\BibitemShut {NoStop}%
\bibitem [{\citenamefont {Elgeti}\ \emph {et~al.}(2015)\citenamefont {Elgeti},
  \citenamefont {Winkler},\ and\ \citenamefont {Gompper}}]{elgeti2015physics}%
  \BibitemOpen
  \bibfield  {author} {\bibinfo {author} {\bibfnamefont {J.}~\bibnamefont
  {Elgeti}}, \bibinfo {author} {\bibfnamefont {R.~G.}\ \bibnamefont
  {Winkler}},\ and\ \bibinfo {author} {\bibfnamefont {G.}~\bibnamefont
  {Gompper}},\ }\href@noop {} {\bibfield  {journal} {\bibinfo  {journal}
  {Reports on progress in physics}\ }\textbf {\bibinfo {volume} {78}},\
  \bibinfo {pages} {056601} (\bibinfo {year} {2015})}\BibitemShut {NoStop}%
\bibitem [{\citenamefont {Qi}\ \emph {et~al.}(2020)\citenamefont {Qi},
  \citenamefont {Westphal}, \citenamefont {Gompper},\ and\ \citenamefont
  {Winkler}}]{qi2020enhanced}%
  \BibitemOpen
  \bibfield  {author} {\bibinfo {author} {\bibfnamefont {K.}~\bibnamefont
  {Qi}}, \bibinfo {author} {\bibfnamefont {E.}~\bibnamefont {Westphal}},
  \bibinfo {author} {\bibfnamefont {G.}~\bibnamefont {Gompper}},\ and\ \bibinfo
  {author} {\bibfnamefont {R.~G.}\ \bibnamefont {Winkler}},\ }\href@noop {}
  {\bibfield  {journal} {\bibinfo  {journal} {Physical Review Letters}\
  }\textbf {\bibinfo {volume} {124}},\ \bibinfo {pages} {068001} (\bibinfo
  {year} {2020})}\BibitemShut {NoStop}%
\bibitem [{\citenamefont {Mousavi}\ \emph {et~al.}(2020)\citenamefont
  {Mousavi}, \citenamefont {Gompper},\ and\ \citenamefont
  {Winkler}}]{mousavi2020wall}%
  \BibitemOpen
  \bibfield  {author} {\bibinfo {author} {\bibfnamefont {S.~M.}\ \bibnamefont
  {Mousavi}}, \bibinfo {author} {\bibfnamefont {G.}~\bibnamefont {Gompper}},\
  and\ \bibinfo {author} {\bibfnamefont {R.~G.}\ \bibnamefont {Winkler}},\
  }\href@noop {} {\bibfield  {journal} {\bibinfo  {journal} {Soft Matter}\
  }\textbf {\bibinfo {volume} {16}},\ \bibinfo {pages} {4866} (\bibinfo {year}
  {2020})}\BibitemShut {NoStop}%
\bibitem [{\citenamefont {Das}\ and\ \citenamefont
  {Chelakkot}(2020)}]{D0SM00797H}%
  \BibitemOpen
  \bibfield  {author} {\bibinfo {author} {\bibfnamefont {S.}~\bibnamefont
  {Das}}\ and\ \bibinfo {author} {\bibfnamefont {R.}~\bibnamefont
  {Chelakkot}},\ }\href@noop {} {\bibfield  {journal} {\bibinfo  {journal}
  {Soft Matter}\ }\textbf {\bibinfo {volume} {16}},\ \bibinfo {pages} {7250}
  (\bibinfo {year} {2020})}\BibitemShut {NoStop}%
\bibitem [{\citenamefont {Patteson}\ \emph {et~al.}(2018)\citenamefont
  {Patteson}, \citenamefont {Gopinath},\ and\ \citenamefont
  {Arratia}}]{patteson2018propagation}%
  \BibitemOpen
  \bibfield  {author} {\bibinfo {author} {\bibfnamefont {A.~E.}\ \bibnamefont
  {Patteson}}, \bibinfo {author} {\bibfnamefont {A.}~\bibnamefont {Gopinath}},\
  and\ \bibinfo {author} {\bibfnamefont {P.~E.}\ \bibnamefont {Arratia}},\
  }\href@noop {} {\bibfield  {journal} {\bibinfo  {journal} {Nature
  communications}\ }\textbf {\bibinfo {volume} {9}},\ \bibinfo {pages} {1}
  (\bibinfo {year} {2018})}\BibitemShut {NoStop}%
\bibitem [{\citenamefont {Theeyancheri}\ \emph {et~al.}(2020)\citenamefont
  {Theeyancheri}, \citenamefont {Chaki}, \citenamefont {Samanta}, \citenamefont
  {Goswami}, \citenamefont {Chelakkot},\ and\ \citenamefont
  {Chakrabarti}}]{theeyancheri2020translational}%
  \BibitemOpen
  \bibfield  {author} {\bibinfo {author} {\bibfnamefont {L.}~\bibnamefont
  {Theeyancheri}}, \bibinfo {author} {\bibfnamefont {S.}~\bibnamefont {Chaki}},
  \bibinfo {author} {\bibfnamefont {N.}~\bibnamefont {Samanta}}, \bibinfo
  {author} {\bibfnamefont {R.}~\bibnamefont {Goswami}}, \bibinfo {author}
  {\bibfnamefont {R.}~\bibnamefont {Chelakkot}},\ and\ \bibinfo {author}
  {\bibfnamefont {R.}~\bibnamefont {Chakrabarti}},\ }\href@noop {} {\bibfield
  {journal} {\bibinfo  {journal} {Soft Matter}\ }\textbf {\bibinfo {volume}
  {16}},\ \bibinfo {pages} {8482} (\bibinfo {year} {2020})}\BibitemShut
  {NoStop}%
\bibitem [{\citenamefont {Das}\ \emph {et~al.}(2020)\citenamefont {Das},
  \citenamefont {Ghosh},\ and\ \citenamefont {Chelakkot}}]{das2020aggregate}%
  \BibitemOpen
  \bibfield  {author} {\bibinfo {author} {\bibfnamefont {S.}~\bibnamefont
  {Das}}, \bibinfo {author} {\bibfnamefont {S.}~\bibnamefont {Ghosh}},\ and\
  \bibinfo {author} {\bibfnamefont {R.}~\bibnamefont {Chelakkot}},\ }\href@noop
  {} {\bibfield  {journal} {\bibinfo  {journal} {Physical Review E}\ }\textbf
  {\bibinfo {volume} {102}},\ \bibinfo {pages} {032619} (\bibinfo {year}
  {2020})}\BibitemShut {NoStop}%
\bibitem [{\citenamefont {Attygalle}\ and\ \citenamefont
  {Morgan}(1985)}]{attygalle1985ant}%
  \BibitemOpen
  \bibfield  {author} {\bibinfo {author} {\bibfnamefont {A.~B.}\ \bibnamefont
  {Attygalle}}\ and\ \bibinfo {author} {\bibfnamefont {E.~D.}\ \bibnamefont
  {Morgan}},\ }in\ \href@noop {} {\emph {\bibinfo {booktitle} {Advances in
  insect physiology}}},\ Vol.~\bibinfo {volume} {18}\ (\bibinfo  {publisher}
  {Elsevier},\ \bibinfo {year} {1985})\ pp.\ \bibinfo {pages}
  {1--30}\BibitemShut {NoStop}%
\bibitem [{\citenamefont {Jackson}\ \emph {et~al.}(2004)\citenamefont
  {Jackson}, \citenamefont {Holcombe},\ and\ \citenamefont
  {Ratnieks}}]{jackson2004trail}%
  \BibitemOpen
  \bibfield  {author} {\bibinfo {author} {\bibfnamefont {D.~E.}\ \bibnamefont
  {Jackson}}, \bibinfo {author} {\bibfnamefont {M.}~\bibnamefont {Holcombe}},\
  and\ \bibinfo {author} {\bibfnamefont {F.~L.}\ \bibnamefont {Ratnieks}},\
  }\href@noop {} {\bibfield  {journal} {\bibinfo  {journal} {Nature}\ }\textbf
  {\bibinfo {volume} {432}},\ \bibinfo {pages} {907} (\bibinfo {year}
  {2004})}\BibitemShut {NoStop}%
\bibitem [{\citenamefont {Gloag}\ \emph {et~al.}(2013)\citenamefont {Gloag},
  \citenamefont {Turnbull}, \citenamefont {Huang}, \citenamefont {Vallotton},
  \citenamefont {Wang}, \citenamefont {Nolan}, \citenamefont {Mililli},
  \citenamefont {Hunt}, \citenamefont {Lu}, \citenamefont {Osvath} \emph
  {et~al.}}]{gloag2013self}%
  \BibitemOpen
  \bibfield  {author} {\bibinfo {author} {\bibfnamefont {E.~S.}\ \bibnamefont
  {Gloag}}, \bibinfo {author} {\bibfnamefont {L.}~\bibnamefont {Turnbull}},
  \bibinfo {author} {\bibfnamefont {A.}~\bibnamefont {Huang}}, \bibinfo
  {author} {\bibfnamefont {P.}~\bibnamefont {Vallotton}}, \bibinfo {author}
  {\bibfnamefont {H.}~\bibnamefont {Wang}}, \bibinfo {author} {\bibfnamefont
  {L.~M.}\ \bibnamefont {Nolan}}, \bibinfo {author} {\bibfnamefont
  {L.}~\bibnamefont {Mililli}}, \bibinfo {author} {\bibfnamefont
  {C.}~\bibnamefont {Hunt}}, \bibinfo {author} {\bibfnamefont {J.}~\bibnamefont
  {Lu}}, \bibinfo {author} {\bibfnamefont {S.~R.}\ \bibnamefont {Osvath}},
  \emph {et~al.},\ }\href@noop {} {\bibfield  {journal} {\bibinfo  {journal}
  {Proceedings of the National Academy of Sciences}\ }\textbf {\bibinfo
  {volume} {110}},\ \bibinfo {pages} {11541} (\bibinfo {year}
  {2013})}\BibitemShut {NoStop}%
\bibitem [{\citenamefont {Gloag}\ \emph {et~al.}(2015)\citenamefont {Gloag},
  \citenamefont {Turnbull},\ and\ \citenamefont
  {Whitchurch}}]{gloag2015bacterial}%
  \BibitemOpen
  \bibfield  {author} {\bibinfo {author} {\bibfnamefont {E.~S.}\ \bibnamefont
  {Gloag}}, \bibinfo {author} {\bibfnamefont {L.}~\bibnamefont {Turnbull}},\
  and\ \bibinfo {author} {\bibfnamefont {C.~B.}\ \bibnamefont {Whitchurch}},\
  }\href@noop {} {\bibfield  {journal} {\bibinfo  {journal} {Scientifica}\
  }\textbf {\bibinfo {volume} {2015}} (\bibinfo {year} {2015})}\BibitemShut
  {NoStop}%
\bibitem [{\citenamefont {Gloag}\ \emph {et~al.}(2016)\citenamefont {Gloag},
  \citenamefont {Turnbull}, \citenamefont {Javed}, \citenamefont {Wang},
  \citenamefont {Gee}, \citenamefont {Wade},\ and\ \citenamefont
  {Whitchurch}}]{gloag2016stigmergy}%
  \BibitemOpen
  \bibfield  {author} {\bibinfo {author} {\bibfnamefont {E.~S.}\ \bibnamefont
  {Gloag}}, \bibinfo {author} {\bibfnamefont {L.}~\bibnamefont {Turnbull}},
  \bibinfo {author} {\bibfnamefont {M.~A.}\ \bibnamefont {Javed}}, \bibinfo
  {author} {\bibfnamefont {H.}~\bibnamefont {Wang}}, \bibinfo {author}
  {\bibfnamefont {M.~L.}\ \bibnamefont {Gee}}, \bibinfo {author} {\bibfnamefont
  {S.~A.}\ \bibnamefont {Wade}},\ and\ \bibinfo {author} {\bibfnamefont
  {C.~B.}\ \bibnamefont {Whitchurch}},\ }\href@noop {} {\bibfield  {journal}
  {\bibinfo  {journal} {Scientific reports}\ }\textbf {\bibinfo {volume} {6}},\
  \bibinfo {pages} {26005} (\bibinfo {year} {2016})}\BibitemShut {NoStop}%
\bibitem [{\citenamefont {Farrell}\ \emph {et~al.}(2013)\citenamefont
  {Farrell}, \citenamefont {Hallatschek}, \citenamefont {Marenmanzzo},\ and\
  \citenamefont {Waclaw}}]{farrell2013mechanically}%
  \BibitemOpen
  \bibfield  {author} {\bibinfo {author} {\bibfnamefont {F.}~\bibnamefont
  {Farrell}}, \bibinfo {author} {\bibfnamefont {O.}~\bibnamefont
  {Hallatschek}}, \bibinfo {author} {\bibfnamefont {D.}~\bibnamefont
  {Marenmanzzo}},\ and\ \bibinfo {author} {\bibfnamefont {B.}~\bibnamefont
  {Waclaw}},\ }\href@noop {} {\bibfield  {journal} {\bibinfo  {journal}
  {Physical review letters}\ }\textbf {\bibinfo {volume} {111}},\ \bibinfo
  {pages} {168101} (\bibinfo {year} {2013})}\BibitemShut {NoStop}%
\bibitem [{\citenamefont {Ghosh}\ \emph {et~al.}(2015)\citenamefont {Ghosh},
  \citenamefont {Mondal}, \citenamefont {Ben-Jacob},\ and\ \citenamefont
  {Levine}}]{ghosh2015mechanically}%
  \BibitemOpen
  \bibfield  {author} {\bibinfo {author} {\bibfnamefont {P.}~\bibnamefont
  {Ghosh}}, \bibinfo {author} {\bibfnamefont {J.}~\bibnamefont {Mondal}},
  \bibinfo {author} {\bibfnamefont {E.}~\bibnamefont {Ben-Jacob}},\ and\
  \bibinfo {author} {\bibfnamefont {H.}~\bibnamefont {Levine}},\ }\href@noop {}
  {\bibfield  {journal} {\bibinfo  {journal} {Proceedings of the National
  Academy of Sciences}\ }\textbf {\bibinfo {volume} {112}},\ \bibinfo {pages}
  {E2166} (\bibinfo {year} {2015})}\BibitemShut {NoStop}%
\bibitem [{\citenamefont {Tchoufag}\ \emph {et~al.}(2019)\citenamefont
  {Tchoufag}, \citenamefont {Ghosh}, \citenamefont {Pogue}, \citenamefont
  {Nan},\ and\ \citenamefont {Mandadapu}}]{tchoufag2019mechanisms}%
  \BibitemOpen
  \bibfield  {author} {\bibinfo {author} {\bibfnamefont {J.}~\bibnamefont
  {Tchoufag}}, \bibinfo {author} {\bibfnamefont {P.}~\bibnamefont {Ghosh}},
  \bibinfo {author} {\bibfnamefont {C.~B.}\ \bibnamefont {Pogue}}, \bibinfo
  {author} {\bibfnamefont {B.}~\bibnamefont {Nan}},\ and\ \bibinfo {author}
  {\bibfnamefont {K.~K.}\ \bibnamefont {Mandadapu}},\ }\href@noop {} {\bibfield
   {journal} {\bibinfo  {journal} {Proceedings of the National Academy of
  Sciences}\ }\textbf {\bibinfo {volume} {116}},\ \bibinfo {pages} {25087}
  (\bibinfo {year} {2019})}\BibitemShut {NoStop}%
\bibitem [{\citenamefont {You}\ \emph {et~al.}(2018)\citenamefont {You},
  \citenamefont {Pearce}, \citenamefont {Sengupta},\ and\ \citenamefont
  {Giomi}}]{you2018geometry}%
  \BibitemOpen
  \bibfield  {author} {\bibinfo {author} {\bibfnamefont {Z.}~\bibnamefont
  {You}}, \bibinfo {author} {\bibfnamefont {D.~J.}\ \bibnamefont {Pearce}},
  \bibinfo {author} {\bibfnamefont {A.}~\bibnamefont {Sengupta}},\ and\
  \bibinfo {author} {\bibfnamefont {L.}~\bibnamefont {Giomi}},\ }\href@noop {}
  {\bibfield  {journal} {\bibinfo  {journal} {Physical Review X}\ }\textbf
  {\bibinfo {volume} {8}},\ \bibinfo {pages} {031065} (\bibinfo {year}
  {2018})}\BibitemShut {NoStop}%
\bibitem [{\citenamefont {Acemel}\ \emph {et~al.}(2018)\citenamefont {Acemel},
  \citenamefont {Govantes},\ and\ \citenamefont {Cuetos}}]{acemel2018computer}%
  \BibitemOpen
  \bibfield  {author} {\bibinfo {author} {\bibfnamefont {R.~D.}\ \bibnamefont
  {Acemel}}, \bibinfo {author} {\bibfnamefont {F.}~\bibnamefont {Govantes}},\
  and\ \bibinfo {author} {\bibfnamefont {A.}~\bibnamefont {Cuetos}},\
  }\href@noop {} {\bibfield  {journal} {\bibinfo  {journal} {Scientific
  reports}\ }\textbf {\bibinfo {volume} {8}},\ \bibinfo {pages} {1} (\bibinfo
  {year} {2018})}\BibitemShut {NoStop}%
\bibitem [{\citenamefont {Farrell}\ \emph {et~al.}(2017)\citenamefont
  {Farrell}, \citenamefont {Gralka}, \citenamefont {Hallatschek},\ and\
  \citenamefont {Waclaw}}]{farrell2017mechanical}%
  \BibitemOpen
  \bibfield  {author} {\bibinfo {author} {\bibfnamefont {F.~D.}\ \bibnamefont
  {Farrell}}, \bibinfo {author} {\bibfnamefont {M.}~\bibnamefont {Gralka}},
  \bibinfo {author} {\bibfnamefont {O.}~\bibnamefont {Hallatschek}},\ and\
  \bibinfo {author} {\bibfnamefont {B.}~\bibnamefont {Waclaw}},\ }\href@noop {}
  {\bibfield  {journal} {\bibinfo  {journal} {Journal of The Royal Society
  Interface}\ }\textbf {\bibinfo {volume} {14}},\ \bibinfo {pages} {20170073}
  (\bibinfo {year} {2017})}\BibitemShut {NoStop}%
\bibitem [{\citenamefont {You}\ \emph {et~al.}(2019)\citenamefont {You},
  \citenamefont {Pearce}, \citenamefont {Sengupta},\ and\ \citenamefont
  {Giomi}}]{you2019mono}%
  \BibitemOpen
  \bibfield  {author} {\bibinfo {author} {\bibfnamefont {Z.}~\bibnamefont
  {You}}, \bibinfo {author} {\bibfnamefont {D.~J.}\ \bibnamefont {Pearce}},
  \bibinfo {author} {\bibfnamefont {A.}~\bibnamefont {Sengupta}},\ and\
  \bibinfo {author} {\bibfnamefont {L.}~\bibnamefont {Giomi}},\ }\href@noop {}
  {\bibfield  {journal} {\bibinfo  {journal} {Physical review letters}\
  }\textbf {\bibinfo {volume} {123}},\ \bibinfo {pages} {178001} (\bibinfo
  {year} {2019})}\BibitemShut {NoStop}%
\bibitem [{\citenamefont {Peruani}\ \emph {et~al.}(2006)\citenamefont
  {Peruani}, \citenamefont {Deutsch},\ and\ \citenamefont
  {B{\"a}r}}]{peruani2006nonequilibrium}%
  \BibitemOpen
  \bibfield  {author} {\bibinfo {author} {\bibfnamefont {F.}~\bibnamefont
  {Peruani}}, \bibinfo {author} {\bibfnamefont {A.}~\bibnamefont {Deutsch}},\
  and\ \bibinfo {author} {\bibfnamefont {M.}~\bibnamefont {B{\"a}r}},\
  }\href@noop {} {\bibfield  {journal} {\bibinfo  {journal} {Physical Review
  E}\ }\textbf {\bibinfo {volume} {74}},\ \bibinfo {pages} {030904} (\bibinfo
  {year} {2006})}\BibitemShut {NoStop}%
\bibitem [{\citenamefont {McCandlish}\ \emph {et~al.}(2012)\citenamefont
  {McCandlish}, \citenamefont {Baskaran},\ and\ \citenamefont
  {Hagan}}]{mccandlish2012spontaneous}%
  \BibitemOpen
  \bibfield  {author} {\bibinfo {author} {\bibfnamefont {S.~R.}\ \bibnamefont
  {McCandlish}}, \bibinfo {author} {\bibfnamefont {A.}~\bibnamefont
  {Baskaran}},\ and\ \bibinfo {author} {\bibfnamefont {M.~F.}\ \bibnamefont
  {Hagan}},\ }\href@noop {} {\bibfield  {journal} {\bibinfo  {journal} {Soft
  Matter}\ }\textbf {\bibinfo {volume} {8}},\ \bibinfo {pages} {2527} (\bibinfo
  {year} {2012})}\BibitemShut {NoStop}%
\bibitem [{\citenamefont {Abkenar}\ \emph {et~al.}(2013)\citenamefont
  {Abkenar}, \citenamefont {Marx}, \citenamefont {Auth},\ and\ \citenamefont
  {Gompper}}]{abkenar2013collective}%
  \BibitemOpen
  \bibfield  {author} {\bibinfo {author} {\bibfnamefont {M.}~\bibnamefont
  {Abkenar}}, \bibinfo {author} {\bibfnamefont {K.}~\bibnamefont {Marx}},
  \bibinfo {author} {\bibfnamefont {T.}~\bibnamefont {Auth}},\ and\ \bibinfo
  {author} {\bibfnamefont {G.}~\bibnamefont {Gompper}},\ }\href@noop {}
  {\bibfield  {journal} {\bibinfo  {journal} {Physical Review E}\ }\textbf
  {\bibinfo {volume} {88}},\ \bibinfo {pages} {062314} (\bibinfo {year}
  {2013})}\BibitemShut {NoStop}%
\bibitem [{\citenamefont {Baskaran}\ and\ \citenamefont
  {Marchetti}(2008)}]{baskaran2008hydrodynamics}%
  \BibitemOpen
  \bibfield  {author} {\bibinfo {author} {\bibfnamefont {A.}~\bibnamefont
  {Baskaran}}\ and\ \bibinfo {author} {\bibfnamefont {M.~C.}\ \bibnamefont
  {Marchetti}},\ }\href@noop {} {\bibfield  {journal} {\bibinfo  {journal}
  {Physical Review E}\ }\textbf {\bibinfo {volume} {77}},\ \bibinfo {pages}
  {011920} (\bibinfo {year} {2008})}\BibitemShut {NoStop}%
\bibitem [{\citenamefont {Weitz}\ \emph {et~al.}(2015)\citenamefont {Weitz},
  \citenamefont {Deutsch},\ and\ \citenamefont {Peruani}}]{weitz2015self}%
  \BibitemOpen
  \bibfield  {author} {\bibinfo {author} {\bibfnamefont {S.}~\bibnamefont
  {Weitz}}, \bibinfo {author} {\bibfnamefont {A.}~\bibnamefont {Deutsch}},\
  and\ \bibinfo {author} {\bibfnamefont {F.}~\bibnamefont {Peruani}},\
  }\href@noop {} {\bibfield  {journal} {\bibinfo  {journal} {Physical Review
  E}\ }\textbf {\bibinfo {volume} {92}},\ \bibinfo {pages} {012322} (\bibinfo
  {year} {2015})}\BibitemShut {NoStop}%
\bibitem [{\citenamefont {Prathyusha}\ \emph {et~al.}(2018)\citenamefont
  {Prathyusha}, \citenamefont {Henkes},\ and\ \citenamefont
  {Sknepnek}}]{prathyusha2018dynamically}%
  \BibitemOpen
  \bibfield  {author} {\bibinfo {author} {\bibfnamefont {K.}~\bibnamefont
  {Prathyusha}}, \bibinfo {author} {\bibfnamefont {S.}~\bibnamefont {Henkes}},\
  and\ \bibinfo {author} {\bibfnamefont {R.}~\bibnamefont {Sknepnek}},\
  }\href@noop {} {\bibfield  {journal} {\bibinfo  {journal} {Physical Review
  E}\ }\textbf {\bibinfo {volume} {97}},\ \bibinfo {pages} {022606} (\bibinfo
  {year} {2018})}\BibitemShut {NoStop}%
\bibitem [{\citenamefont {Velasco}\ \emph {et~al.}(2018)\citenamefont
  {Velasco}, \citenamefont {Abkenar}, \citenamefont {Gompper},\ and\
  \citenamefont {Auth}}]{velasco2018collective}%
  \BibitemOpen
  \bibfield  {author} {\bibinfo {author} {\bibfnamefont {C.~A.}\ \bibnamefont
  {Velasco}}, \bibinfo {author} {\bibfnamefont {M.}~\bibnamefont {Abkenar}},
  \bibinfo {author} {\bibfnamefont {G.}~\bibnamefont {Gompper}},\ and\ \bibinfo
  {author} {\bibfnamefont {T.}~\bibnamefont {Auth}},\ }\href@noop {} {\bibfield
   {journal} {\bibinfo  {journal} {Physical Review E}\ }\textbf {\bibinfo
  {volume} {98}},\ \bibinfo {pages} {022605} (\bibinfo {year}
  {2018})}\BibitemShut {NoStop}%
\bibitem [{\citenamefont {Kuan}\ \emph {et~al.}(2015)\citenamefont {Kuan},
  \citenamefont {Blackwell}, \citenamefont {Hough}, \citenamefont {Glaser},\
  and\ \citenamefont {Betterton}}]{kuan2015hysteresis}%
  \BibitemOpen
  \bibfield  {author} {\bibinfo {author} {\bibfnamefont {H.-S.}\ \bibnamefont
  {Kuan}}, \bibinfo {author} {\bibfnamefont {R.}~\bibnamefont {Blackwell}},
  \bibinfo {author} {\bibfnamefont {L.~E.}\ \bibnamefont {Hough}}, \bibinfo
  {author} {\bibfnamefont {M.~A.}\ \bibnamefont {Glaser}},\ and\ \bibinfo
  {author} {\bibfnamefont {M.~D.}\ \bibnamefont {Betterton}},\ }\href@noop {}
  {\bibfield  {journal} {\bibinfo  {journal} {Physical Review E}\ }\textbf
  {\bibinfo {volume} {92}},\ \bibinfo {pages} {060501} (\bibinfo {year}
  {2015})}\BibitemShut {NoStop}%
\bibitem [{\citenamefont {B{\"a}r}\ \emph {et~al.}(2020)\citenamefont
  {B{\"a}r}, \citenamefont {Gro{\ss}mann}, \citenamefont {Heidenreich},\ and\
  \citenamefont {Peruani}}]{bar2020self}%
  \BibitemOpen
  \bibfield  {author} {\bibinfo {author} {\bibfnamefont {M.}~\bibnamefont
  {B{\"a}r}}, \bibinfo {author} {\bibfnamefont {R.}~\bibnamefont
  {Gro{\ss}mann}}, \bibinfo {author} {\bibfnamefont {S.}~\bibnamefont
  {Heidenreich}},\ and\ \bibinfo {author} {\bibfnamefont {F.}~\bibnamefont
  {Peruani}},\ }\href@noop {} {\bibfield  {journal} {\bibinfo  {journal}
  {Annual Review of Condensed Matter Physics}\ }\textbf {\bibinfo {volume}
  {11}},\ \bibinfo {pages} {441} (\bibinfo {year} {2020})}\BibitemShut
  {NoStop}%
\bibitem [{\citenamefont {Be’er}\ \emph {et~al.}(2020)\citenamefont
  {Be’er}, \citenamefont {Ilkanaiv}, \citenamefont {Gross}, \citenamefont
  {Kearns}, \citenamefont {Heidenreich}, \citenamefont {B{\"a}r},\ and\
  \citenamefont {Ariel}}]{be2020phase}%
  \BibitemOpen
  \bibfield  {author} {\bibinfo {author} {\bibfnamefont {A.}~\bibnamefont
  {Be’er}}, \bibinfo {author} {\bibfnamefont {B.}~\bibnamefont {Ilkanaiv}},
  \bibinfo {author} {\bibfnamefont {R.}~\bibnamefont {Gross}}, \bibinfo
  {author} {\bibfnamefont {D.~B.}\ \bibnamefont {Kearns}}, \bibinfo {author}
  {\bibfnamefont {S.}~\bibnamefont {Heidenreich}}, \bibinfo {author}
  {\bibfnamefont {M.}~\bibnamefont {B{\"a}r}},\ and\ \bibinfo {author}
  {\bibfnamefont {G.}~\bibnamefont {Ariel}},\ }\href@noop {} {\bibfield
  {journal} {\bibinfo  {journal} {Communications Physics}\ }\textbf {\bibinfo
  {volume} {3}},\ \bibinfo {pages} {1} (\bibinfo {year} {2020})}\BibitemShut
  {NoStop}%
\bibitem [{\citenamefont {Vliegenthart}\ \emph {et~al.}(2020)\citenamefont
  {Vliegenthart}, \citenamefont {Ravichandran}, \citenamefont {Ripoll},
  \citenamefont {Auth},\ and\ \citenamefont
  {Gompper}}]{vliegenthart2020filamentous}%
  \BibitemOpen
  \bibfield  {author} {\bibinfo {author} {\bibfnamefont {G.~A.}\ \bibnamefont
  {Vliegenthart}}, \bibinfo {author} {\bibfnamefont {A.}~\bibnamefont
  {Ravichandran}}, \bibinfo {author} {\bibfnamefont {M.}~\bibnamefont
  {Ripoll}}, \bibinfo {author} {\bibfnamefont {T.}~\bibnamefont {Auth}},\ and\
  \bibinfo {author} {\bibfnamefont {G.}~\bibnamefont {Gompper}},\ }\href@noop
  {} {\bibfield  {journal} {\bibinfo  {journal} {Science advances}\ }\textbf
  {\bibinfo {volume} {6}},\ \bibinfo {pages} {eaaw9975} (\bibinfo {year}
  {2020})}\BibitemShut {NoStop}%
\bibitem [{\citenamefont {Bott}\ \emph {et~al.}(2018)\citenamefont {Bott},
  \citenamefont {Winterhalter}, \citenamefont {Marechal}, \citenamefont
  {Sharma}, \citenamefont {Brader},\ and\ \citenamefont
  {Wittmann}}]{bott2018isotropic}%
  \BibitemOpen
  \bibfield  {author} {\bibinfo {author} {\bibfnamefont {M.~C.}\ \bibnamefont
  {Bott}}, \bibinfo {author} {\bibfnamefont {F.}~\bibnamefont {Winterhalter}},
  \bibinfo {author} {\bibfnamefont {M.}~\bibnamefont {Marechal}}, \bibinfo
  {author} {\bibfnamefont {A.}~\bibnamefont {Sharma}}, \bibinfo {author}
  {\bibfnamefont {J.~M.}\ \bibnamefont {Brader}},\ and\ \bibinfo {author}
  {\bibfnamefont {R.}~\bibnamefont {Wittmann}},\ }\href@noop {} {\bibfield
  {journal} {\bibinfo  {journal} {Physical Review E}\ }\textbf {\bibinfo
  {volume} {98}},\ \bibinfo {pages} {012601} (\bibinfo {year}
  {2018})}\BibitemShut {NoStop}%
\bibitem [{\citenamefont {Zachreson}\ \emph
  {et~al.}(2017{\natexlab{a}})\citenamefont {Zachreson}, \citenamefont {Wolff},
  \citenamefont {Whitchurch},\ and\ \citenamefont
  {Toth}}]{zachreson2017emergent}%
  \BibitemOpen
  \bibfield  {author} {\bibinfo {author} {\bibfnamefont {C.}~\bibnamefont
  {Zachreson}}, \bibinfo {author} {\bibfnamefont {C.}~\bibnamefont {Wolff}},
  \bibinfo {author} {\bibfnamefont {C.~B.}\ \bibnamefont {Whitchurch}},\ and\
  \bibinfo {author} {\bibfnamefont {M.}~\bibnamefont {Toth}},\ }\href@noop {}
  {\bibfield  {journal} {\bibinfo  {journal} {Physical Review E}\ }\textbf
  {\bibinfo {volume} {95}},\ \bibinfo {pages} {012408} (\bibinfo {year}
  {2017}{\natexlab{a}})}\BibitemShut {NoStop}%
\bibitem [{\citenamefont {Zachreson}\ \emph
  {et~al.}(2017{\natexlab{b}})\citenamefont {Zachreson}, \citenamefont {Yap},
  \citenamefont {Gloag}, \citenamefont {Shimoni}, \citenamefont {Whitchurch},\
  and\ \citenamefont {Toth}}]{zachreson2017network}%
  \BibitemOpen
  \bibfield  {author} {\bibinfo {author} {\bibfnamefont {C.}~\bibnamefont
  {Zachreson}}, \bibinfo {author} {\bibfnamefont {X.}~\bibnamefont {Yap}},
  \bibinfo {author} {\bibfnamefont {E.~S.}\ \bibnamefont {Gloag}}, \bibinfo
  {author} {\bibfnamefont {R.}~\bibnamefont {Shimoni}}, \bibinfo {author}
  {\bibfnamefont {C.~B.}\ \bibnamefont {Whitchurch}},\ and\ \bibinfo {author}
  {\bibfnamefont {M.}~\bibnamefont {Toth}},\ }\href@noop {} {\bibfield
  {journal} {\bibinfo  {journal} {Physical Review E}\ }\textbf {\bibinfo
  {volume} {96}},\ \bibinfo {pages} {042401} (\bibinfo {year}
  {2017}{\natexlab{b}})}\BibitemShut {NoStop}%
\bibitem [{\citenamefont {Imaran}\ \emph {et~al.}(2020)\citenamefont {Imaran},
  \citenamefont {Prabhakar}, \citenamefont {Chelakkot},\ and\ \citenamefont
  {Inamdar}}]{SI}%
  \BibitemOpen
  \bibfield  {author} {\bibinfo {author} {\bibfnamefont {M.}~\bibnamefont
  {Imaran}}, \bibinfo {author} {\bibfnamefont {R.}~\bibnamefont {Prabhakar}},
  \bibinfo {author} {\bibfnamefont {R.}~\bibnamefont {Chelakkot}},\ and\
  \bibinfo {author} {\bibfnamefont {M.}~\bibnamefont {Inamdar}},\ }\href@noop
  {} {\  (\bibinfo {year} {2020})}\BibitemShut {NoStop}%
\bibitem [{\citenamefont {Gagnepain}\ and\ \citenamefont
  {Roques-Carmes}(1986)}]{gagnepain1986fractal}%
  \BibitemOpen
  \bibfield  {author} {\bibinfo {author} {\bibfnamefont {J.}~\bibnamefont
  {Gagnepain}}\ and\ \bibinfo {author} {\bibfnamefont {C.}~\bibnamefont
  {Roques-Carmes}},\ }\href@noop {} {\bibfield  {journal} {\bibinfo  {journal}
  {wear}\ }\textbf {\bibinfo {volume} {109}},\ \bibinfo {pages} {119} (\bibinfo
  {year} {1986})}\BibitemShut {NoStop}%
\bibitem [{\citenamefont {Florio}\ \emph {et~al.}(2019)\citenamefont {Florio},
  \citenamefont {Fawell},\ and\ \citenamefont {Small}}]{florio2019use}%
  \BibitemOpen
  \bibfield  {author} {\bibinfo {author} {\bibfnamefont {B.~J.}\ \bibnamefont
  {Florio}}, \bibinfo {author} {\bibfnamefont {P.~D.}\ \bibnamefont {Fawell}},\
  and\ \bibinfo {author} {\bibfnamefont {M.}~\bibnamefont {Small}},\
  }\href@noop {} {\bibfield  {journal} {\bibinfo  {journal} {Powder
  technology}\ }\textbf {\bibinfo {volume} {343}},\ \bibinfo {pages} {551}
  (\bibinfo {year} {2019})}\BibitemShut {NoStop}%
\bibitem [{\citenamefont {Devaraj}\ \emph {et~al.}(2019)\citenamefont
  {Devaraj}, \citenamefont {Buzzo}, \citenamefont {Mashburn-Warren},
  \citenamefont {Gloag}, \citenamefont {Novotny}, \citenamefont {Stoodley},
  \citenamefont {Bakaletz},\ and\ \citenamefont {Goodman}}]{Devaraj2019}%
  \BibitemOpen
  \bibfield  {author} {\bibinfo {author} {\bibfnamefont {A.}~\bibnamefont
  {Devaraj}}, \bibinfo {author} {\bibfnamefont {J.~R.}\ \bibnamefont {Buzzo}},
  \bibinfo {author} {\bibfnamefont {L.}~\bibnamefont {Mashburn-Warren}},
  \bibinfo {author} {\bibfnamefont {E.~S.}\ \bibnamefont {Gloag}}, \bibinfo
  {author} {\bibfnamefont {L.~A.}\ \bibnamefont {Novotny}}, \bibinfo {author}
  {\bibfnamefont {P.}~\bibnamefont {Stoodley}}, \bibinfo {author}
  {\bibfnamefont {L.~O.}\ \bibnamefont {Bakaletz}},\ and\ \bibinfo {author}
  {\bibfnamefont {S.~D.}\ \bibnamefont {Goodman}},\ }\href
  {https://doi.org/10.1073/pnas.1909017116} {\bibfield  {journal} {\bibinfo
  {journal} {Proceedings of the National Academy of Sciences}\ }\textbf
  {\bibinfo {volume} {116}},\ \bibinfo {pages} {25068} (\bibinfo {year}
  {2019})}\BibitemShut {NoStop}%
\bibitem [{\citenamefont {Nagilla}\ \emph {et~al.}(2018)\citenamefont
  {Nagilla}, \citenamefont {Prabhakar},\ and\ \citenamefont
  {Jadhav}}]{Nagilla2018}%
  \BibitemOpen
  \bibfield  {author} {\bibinfo {author} {\bibfnamefont {A.}~\bibnamefont
  {Nagilla}}, \bibinfo {author} {\bibfnamefont {R.}~\bibnamefont {Prabhakar}},\
  and\ \bibinfo {author} {\bibfnamefont {S.}~\bibnamefont {Jadhav}},\
  }\href@noop {} {\bibfield  {journal} {\bibinfo  {journal} {Phys. Fluids}\
  }\textbf {\bibinfo {volume} {30}},\ \bibinfo {pages} {022109} (\bibinfo
  {year} {2018})}\BibitemShut {NoStop}%
\bibitem [{\citenamefont {Giverso}\ \emph {et~al.}(2015)\citenamefont
  {Giverso}, \citenamefont {Verani},\ and\ \citenamefont
  {Ciarletta}}]{giverso2015branching}%
  \BibitemOpen
  \bibfield  {author} {\bibinfo {author} {\bibfnamefont {C.}~\bibnamefont
  {Giverso}}, \bibinfo {author} {\bibfnamefont {M.}~\bibnamefont {Verani}},\
  and\ \bibinfo {author} {\bibfnamefont {P.}~\bibnamefont {Ciarletta}},\
  }\href@noop {} {\bibfield  {journal} {\bibinfo  {journal} {Journal of The
  Royal Society Interface}\ }\textbf {\bibinfo {volume} {12}},\ \bibinfo
  {pages} {20141290} (\bibinfo {year} {2015})}\BibitemShut {NoStop}%
\bibitem [{\citenamefont {Turnbull}\ \emph {et~al.}(2016)\citenamefont
  {Turnbull}, \citenamefont {Toyofuku}, \citenamefont {Hynen}, \citenamefont
  {Kurosawa}, \citenamefont {Pessi}, \citenamefont {Petty}, \citenamefont
  {Osvath}, \citenamefont {C{\'a}rcamo-Oyarce}, \citenamefont {Gloag},
  \citenamefont {Shimoni}, \citenamefont {Omasits}, \citenamefont {Ito},
  \citenamefont {Yap}, \citenamefont {Monahan}, \citenamefont {Cavaliere},
  \citenamefont {Ahrens}, \citenamefont {Charles}, \citenamefont {Nomura},
  \citenamefont {Eberl},\ and\ \citenamefont {Whitchurch}}]{Turnbull2016}%
  \BibitemOpen
  \bibfield  {author} {\bibinfo {author} {\bibfnamefont {L.}~\bibnamefont
  {Turnbull}}, \bibinfo {author} {\bibfnamefont {M.}~\bibnamefont {Toyofuku}},
  \bibinfo {author} {\bibfnamefont {A.~L.}\ \bibnamefont {Hynen}}, \bibinfo
  {author} {\bibfnamefont {M.}~\bibnamefont {Kurosawa}}, \bibinfo {author}
  {\bibfnamefont {G.}~\bibnamefont {Pessi}}, \bibinfo {author} {\bibfnamefont
  {N.~K.}\ \bibnamefont {Petty}}, \bibinfo {author} {\bibfnamefont {S.~R.}\
  \bibnamefont {Osvath}}, \bibinfo {author} {\bibfnamefont {G.}~\bibnamefont
  {C{\'a}rcamo-Oyarce}}, \bibinfo {author} {\bibfnamefont {E.~S.}\ \bibnamefont
  {Gloag}}, \bibinfo {author} {\bibfnamefont {R.}~\bibnamefont {Shimoni}},
  \bibinfo {author} {\bibfnamefont {U.}~\bibnamefont {Omasits}}, \bibinfo
  {author} {\bibfnamefont {S.}~\bibnamefont {Ito}}, \bibinfo {author}
  {\bibfnamefont {X.}~\bibnamefont {Yap}}, \bibinfo {author} {\bibfnamefont
  {L.~G.}\ \bibnamefont {Monahan}}, \bibinfo {author} {\bibfnamefont
  {R.}~\bibnamefont {Cavaliere}}, \bibinfo {author} {\bibfnamefont {C.~H.}\
  \bibnamefont {Ahrens}}, \bibinfo {author} {\bibfnamefont {I.~G.}\
  \bibnamefont {Charles}}, \bibinfo {author} {\bibfnamefont {N.}~\bibnamefont
  {Nomura}}, \bibinfo {author} {\bibfnamefont {L.}~\bibnamefont {Eberl}},\ and\
  \bibinfo {author} {\bibfnamefont {C.~B.}\ \bibnamefont {Whitchurch}},\
  }\href@noop {} {\bibfield  {journal} {\bibinfo  {journal} {Nat. Commun.}\
  }\textbf {\bibinfo {volume} {7}},\ \bibinfo {pages} {11220} (\bibinfo {year}
  {2016})}\BibitemShut {NoStop}%
\bibitem [{\citenamefont {Lee}\ \emph {et~al.}(2017)\citenamefont {Lee},
  \citenamefont {Konen}, \citenamefont {Wilkinson}, \citenamefont {Marcus},\
  and\ \citenamefont {Jiang}}]{lee2017local}%
  \BibitemOpen
  \bibfield  {author} {\bibinfo {author} {\bibfnamefont {B.}~\bibnamefont
  {Lee}}, \bibinfo {author} {\bibfnamefont {J.}~\bibnamefont {Konen}}, \bibinfo
  {author} {\bibfnamefont {S.}~\bibnamefont {Wilkinson}}, \bibinfo {author}
  {\bibfnamefont {A.~I.}\ \bibnamefont {Marcus}},\ and\ \bibinfo {author}
  {\bibfnamefont {Y.}~\bibnamefont {Jiang}},\ }\href@noop {} {\bibfield
  {journal} {\bibinfo  {journal} {Scientific reports}\ }\textbf {\bibinfo
  {volume} {7}},\ \bibinfo {pages} {39498} (\bibinfo {year}
  {2017})}\BibitemShut {NoStop}%
\bibitem [{\citenamefont {Wisdom}\ \emph {et~al.}(2018)\citenamefont {Wisdom},
  \citenamefont {Adebowale}, \citenamefont {Chang}, \citenamefont {Lee},
  \citenamefont {Nam}, \citenamefont {Desai}, \citenamefont {Rossen},
  \citenamefont {Rafat}, \citenamefont {West}, \citenamefont {Hodgson} \emph
  {et~al.}}]{wisdom2018matrix}%
  \BibitemOpen
  \bibfield  {author} {\bibinfo {author} {\bibfnamefont {K.~M.}\ \bibnamefont
  {Wisdom}}, \bibinfo {author} {\bibfnamefont {K.}~\bibnamefont {Adebowale}},
  \bibinfo {author} {\bibfnamefont {J.}~\bibnamefont {Chang}}, \bibinfo
  {author} {\bibfnamefont {J.~Y.}\ \bibnamefont {Lee}}, \bibinfo {author}
  {\bibfnamefont {S.}~\bibnamefont {Nam}}, \bibinfo {author} {\bibfnamefont
  {R.}~\bibnamefont {Desai}}, \bibinfo {author} {\bibfnamefont {N.~S.}\
  \bibnamefont {Rossen}}, \bibinfo {author} {\bibfnamefont {M.}~\bibnamefont
  {Rafat}}, \bibinfo {author} {\bibfnamefont {R.~B.}\ \bibnamefont {West}},
  \bibinfo {author} {\bibfnamefont {L.}~\bibnamefont {Hodgson}}, \emph
  {et~al.},\ }\href@noop {} {\bibfield  {journal} {\bibinfo  {journal} {Nature
  communications}\ }\textbf {\bibinfo {volume} {9}},\ \bibinfo {pages} {1}
  (\bibinfo {year} {2018})}\BibitemShut {NoStop}%
\bibitem [{\citenamefont {Winkler}\ \emph {et~al.}(2020)\citenamefont
  {Winkler}, \citenamefont {Abisoye-Ogunniyan}, \citenamefont {Metcalf},\ and\
  \citenamefont {Werb}}]{winkler2020concepts}%
  \BibitemOpen
  \bibfield  {author} {\bibinfo {author} {\bibfnamefont {J.}~\bibnamefont
  {Winkler}}, \bibinfo {author} {\bibfnamefont {A.}~\bibnamefont
  {Abisoye-Ogunniyan}}, \bibinfo {author} {\bibfnamefont {K.~J.}\ \bibnamefont
  {Metcalf}},\ and\ \bibinfo {author} {\bibfnamefont {Z.}~\bibnamefont
  {Werb}},\ }\href@noop {} {\bibfield  {journal} {\bibinfo  {journal} {Nature
  communications}\ }\textbf {\bibinfo {volume} {11}},\ \bibinfo {pages} {1}
  (\bibinfo {year} {2020})}\BibitemShut {NoStop}%
\end{thebibliography}

%

\end{document}